\documentstyle[11pt,paspconf,epsf]{article}

\begin{document}

\title{A dichotomy between HSB and LSB galaxies}

\author{Marc Verheijen\altaffilmark{1}}
\affil{National Radio Astronomy Observatory, Socorro, New Mexico}

\author{Brent Tully}
\affil{University of Hawaii, Institute for Astronomy, Honolulu, Hawaii}

\altaffiltext{1}{Jansky Fellow, Array Operations Center}

\begin{abstract} 

A complete sample of spiral galaxies in the Ursa Major cluster is imaged
at various optical wavelengths and in the Near-Infrared K$^\prime$-band. 
HI rotation curves were obtained for all gas rich systems.  The
Near-Infrared surface brightness distribution of disk galaxies turns out
to be bimodal; galaxies avoid a domain around $\mu_0^{\rm
i}$(K$^\prime$)$\approx$18.5 mag/arcsec$^2$.  This bimodality is
particularly striking when only the more isolated, non-interacting
systems are considered.  The Luminosity Function of the HSB family of
galaxies is truncated well above the completion limit while the
Luminosity Function of the LSB family is still sharply rising at our
limiting magnitude.  Near-Infrared mass-to-light ratios suggest that HSB
galaxies are close to a kinematic maximum-disk situation while LSB
galaxies are dark matter dominated at all radii.  Assuming equal
Near-Infrared mass-to-light ratios for both HSB and LSB systems, we find
that the gap in the surface brightness distribution corresponds to a
situation in which the baryonic mass is marginally self-gravitating.  We
finally conclude that the luminosity-line width relation is a
fundamental correlation between the amount and distribution of dark
matter mass and the total luminosity, regardless of how the luminous
mass is distributed within the dark mater halo. 

\end{abstract}


\keywords{galaxies: fundamental parameters, kinematics and dynamics, structure}

\section{Introduction}

The distribution of disk central surface brightnesses is determined with
some confidence in the B-band.  It shows a steep rise at the bright end
and remains roughly flat towards fainter surface brightnesses (McGaugh
{\it et al}, 1995; McGaugh, 1996; de Jong, 1996; Bothun {\it et al},
1997 and references therein).  In general, however, this distribution is
determined using B-band photometry and the observed surface brightnesses
are not or obscurely corrected for the effects of inclination and
internal extinction.  Furthermore, the effects of metallicity and age of
a stellar population on the B-band surface brightness of a galaxy can be
considerable.  It might not be surprising that the observed B-band
surface brightness distribution is flat to some extent since large
galaxy-to-galaxy variations in dust content, age and metallicity will
smooth out any underlying, physically interesting distribution like the
stellar surface density or angular momentum distribution.  Therefore,
"...it is possible that a great deal of our present understanding about
the surface brightness distribution of galaxies is based on a parameter
that is very insensitive to the actual physical characteristics of
galaxies" (Davies, 1990). 

In our talks we presented results from a Near-Infrared imaging survey of
a complete volume limited sample of galaxies in the Ursa Major Cluster. 
In the Near-Infrared, internal extinction is neglegible and the age and
metallicity effects on the luminosity of a stellar population are
minimal.  We show that the distribution of Near-Infrared face-on disk
central surface brightnesses is bimodal; there is a relative lack of
spirals with $\mu_0^{\rm i}$(K$^\prime$)$\approx$18.5$^{\rm m}$. 
Analyzing HI rotation curves shows that this gap in the Near-Infrared
surface brightness distribution may correspond to a dynamical
instability when a disk is marginally self-gravitating. 

\section{The Sample and Observations}

The Ursa Major cluster, located in the Supergalactic Plane, consists of
80 galaxies in a $\sim$80 Mpc$^3$ volume at a distance of 15.5 Mpc. 
This volume is overdense by roughly a factor 10 compared to the average
field.  However, the population of galaxies is comparable to the field. 
It comprises mainly spiral galaxies, a dozen S0 and maybe one elliptical
system.  A complete sample of 62 galaxies, intrinsically brighter then
the SMC, has been identified in the B-band.  Since all galaxies are at
the same distance, there is little question about their relative
luminosities and disk scale lengths.  This cluster is described in
detail by Tully {\it et al} (1996). 

Surface photometry in the B, R and I passbands is obtained for all
galaxies in 11 observing runs between February 1984 and March 1996. 
Photometric K$^\prime$-band images of 69 spiral galaxies in Ursa Major
were obtained in May 1991, February 1992 and March 1993 using a 256$^2$
HgCdTe detector on the 24-inch and 88-inch telescopes of the University
of Hawaii on Mauna Kea.  Sixty galaxies form a nearly complete sample
(only U6628 and U7129 have missing K$^\prime$-band data) of spirals
intrinsically brighter than the SMC. Images and luminosity profiles can
be found in Tully {\it et al} (1996).

HI synthesis data for all galaxies with sufficient HI gas were obtained
with the Westerbork Synthesis Radio Telescope between 1991 and 1996. 
From these data, HI rotation curves were derived by fitting tilted rings
to the HI velocity fields.  The rotation curves were corrected for the
effects of beam smearing by overlaying them on position-velocity
diagrams and adjusting them by eye.  The radial surface density profiles
of the gas were derived from the total HI maps by averaging the flux in
ellipses following the adjusted tilted rings fits.  The HI data are
described in detail by Verheijen (1997). 

\begin{figure} 
\plotone{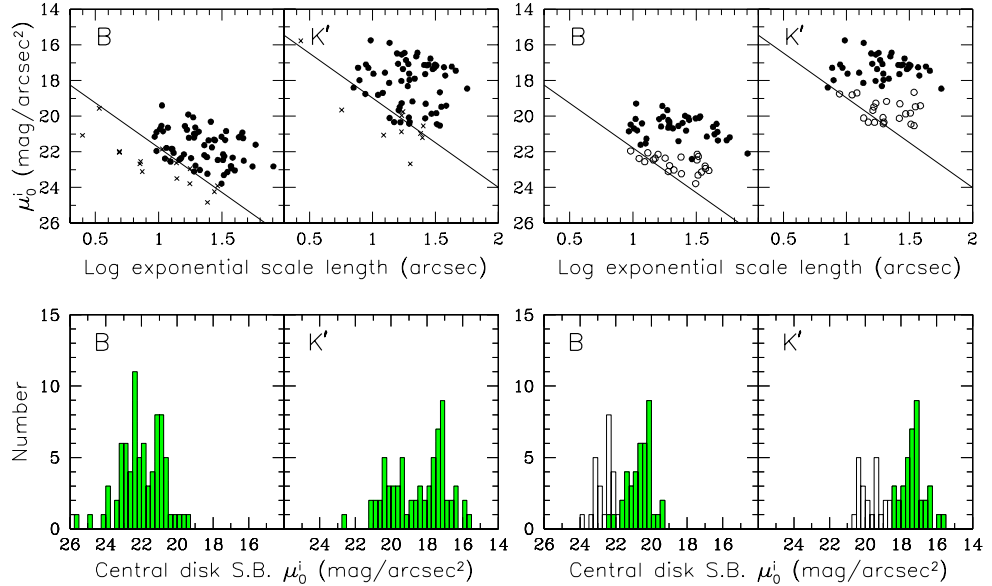} 
\caption{{\bf Upper-left panels:} Plots of central disk surface
brightness versus the log of the exponential disk scale length as
measured in the B and K$^\prime$ bands.  Galaxies are assumed to be
transparent in both passbands ($C^{\rm B,K^\prime}$=1).  Slanted lines
indicate the completion limit for purely exponential disks.  Crosses
indicate galaxies fainter than the SMC.  {\bf Upper-right panels:}
Complete sample only while the LSB family of galaxies is identified with
open symbols.  Here, $C^{\rm K^\prime}_{\rm HSB,LSB}$=1 while $C^{\rm
B}_{\rm HSB}$=0.23 to correct for internal extinction and $C^{\rm
B}_{\rm LSB}$=1.  {\bf Lower panels:} Histogram distributions of the
central disk surface brightnesses corresponding to the upper panels.}
\end{figure}

\begin{figure}
\plotone{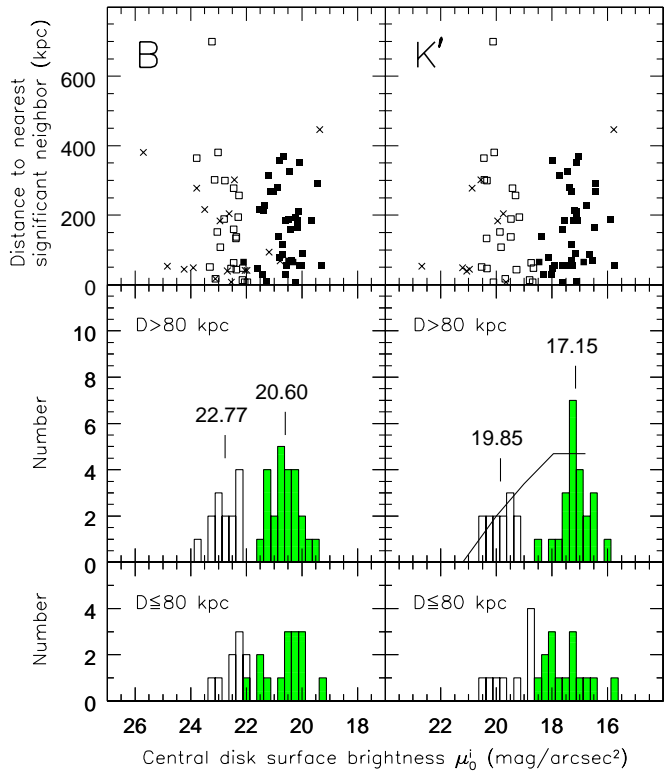}
\caption{{\bf Upper panels:} Disk central surface brightness in the B
and K$^\prime$ passbands versus projected distance to nearest
significant neighbor.  Nearly all galaxies with intermediate surface
brightness have a close companion.  {\bf Middle panels:} Surface
brightness distributions for galaxies more isolated than 80 kpc from
their nearest neighbor.  The line in the K$^\prime$-band panel roughly
indicates the completion limit based on purely exponential disks.  {\bf
Lower panels:} Surface brightness distributions for galaxies with close
companions (in projection).}
\end{figure}

\section{Surface Brightness Distributions}

From the calibrated optical and Near-Infrared images, luminosity
profiles were extracted by averaging the flux density in concentric
ellipses of equal ellipticity and position angle.  Exponential disk fits
were made by fitting straight lines to the quasi-linear part of the
luminosity profiles, yielding disk scale lengths and central surface
brightnesses.  A possible upturn due to a bulge in the inner parts of
the profiles was carefully excluded from the fit.  The following
discusssion of the photometric results will be restricted to the B- and
K$^\prime$-band data. 

\subsection{A Bimodal Distribution}

The measured central surface brightnesses $\mu_0(\lambda)$ at a certain
wavelength $\lambda$ were corrected for inclination and internal
extinction according to $$\mu_0^{\rm i}(\lambda) = \mu_0(\lambda)
-2.5\:C^\lambda\:\mbox{Log(b/a)}$$ were $C^\lambda$ accounts for
internal extinction.  If galaxies are transparent, $C^\lambda$=1 and
$\mu_0(\lambda)$ will only be corrected for the path length through the
inclined disk while $C^\lambda$$<$1 takes a correction for internal
extinction into account. Values of $\mu_0^{\rm i}$ are plotted against disk
scale lengths in Figure 1. LSB galaxies are indicated with open symbols
in the right panels in which a trend of surface brightness with total
luminosity can be seen.

We assume that galaxies are transparent in the Near-Infrared and thus
use $C^{\rm K^\prime}$=1.  Doing so, we find an apparent lack of galaxies
with $\mu_0^{\rm i}$(K$^\prime$)$\approx$18.5 mag/arcsec$^2$ (see
K$^\prime$-panels in Figure 1).  This is a first hint to a possible
bimodality in the Near-Infrared surface brightness distribution.  We use
the Near-Infrared data to define a low surface brightness galaxy in case
its $\mu_0^{\rm i}$(K$^\prime$)$>$18.5 mag/arcsec$^2$ and a high surface
brightness galaxy if $\mu_0^{\rm i}$(K$^\prime$)$<$18.5~mag/arcsec$^2$. 

It should be pointed out that de Jong (1996) obtained K$^\prime$-band
images of 85 galaxies in his sample and derived a Near-Infrared surface
brightness distribution without a hint of any bimodality. 
Unfortunately, his sample is dominated by HSB galaxies and the volume
corrections for the LSB systems are quite uncertain. 

The internal extinction in the B-band gives rise to much more
uncertainty in any correction of $\mu_0$(B) to face-on.  Figure 1 shows
the distribution of $\mu_0^{\rm i}$(B) assuming that either $C^{\rm
B}$=1 for all galaxies (left B-band panels) or $C^{\rm B}$=1 only for
LSB galaxies while $C^{\rm B}$=0.23 for HSB galaxies (right B-band
panels).  The latter assumption is based on the notion that the overall
fainter LSB galaxies in Ursa Major are in general not detected by IRAS,
indicating a low dust content.  Given this bimodal extinction correction
in the B-band, the resulting bimodality of the B-band surface brightness
distribution is largely artificial.  It should be noted that the former
assumption, $C^{\rm B}$=1 for all galaxies, leads to a flat surface
brightness distribution in the B-band. 

\subsection{A Near-Neighbor Effect}

Investigating whether the LSB galaxies in our sample are more isolated
than the HSB galaxies lead to an interesting result.  In the upper
panels of Figure 2 we plot the surface brightness of a galaxy versus the
projected distance to its nearest significant neighbor; the nearest
galaxy with an I-band luminosity of at least 10\% of the galaxy
considered.  It turns out that galaxies of intermediate surface
brightness, $\mu_0^{\rm i}$(K$^\prime$)$\approx$18.5 mag/arcsec$^2$, have
near companions and are often involved in a tidal interaction. 
Considering the more isolated galaxies, the bimodality in the surface
brightness distribution is striking (middle panels).  The line in the
middle K$^\prime$-panel indicates how our completion limit maps onto the
surface brightness distribution.  Within our sample, the LSB galaxies
are not more isolated than the HSB galaxies although the most isolated
system in our sample is an LSB. 

\begin{figure}
\plotone{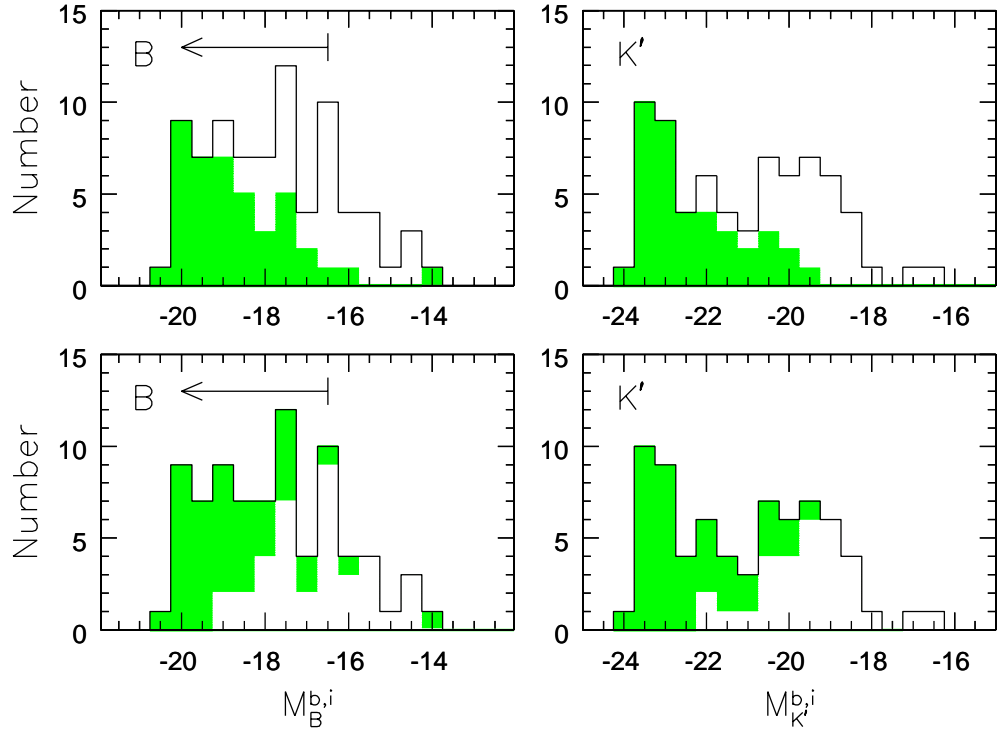}
\caption{Luminosity Functions in the B and K$^\prime$ passbands
including all identified cluster members.  The open part of the
histograms correspond to the LSB family. The horizontal arrow in the
B-band panels indicates the complete sample. The Luminosity Function
of the HSB galaxies drops well above the completion limit. The
sub-histograms are switched between the upper and lower panels. }
\end{figure}

\section{The Luminosity Functions.}

This bimodality of HSB and LSB families is also reflected in the
Luminosity Functions.  Figure 3 shows a linear version of the Luminosity
Functions in both the B and K$^\prime$ bands including galaxies fainter
than the completion limit.  The filled part of the histogram corresponds
to the HSB family of galaxies.  The Luminosity Functions of the HSB
galaxies clearly cut off well above the completion limit while the
Luminosity Functions of the LSB galaxies keep rising toward the faint
end until the completion limit is reached.  In the K$^\prime$-band, the
HSB and LSB Luminosity Functions are more separated than in the B-band. 
This may result in a shallow dip in the Luminosity Functions around
M$_{\rm K^\prime}^{\rm b,i}$=$-$21.5$^{\rm m}$ while the B-band Luminosity Functions is
quite flat. 

\begin{figure} 
\plotone{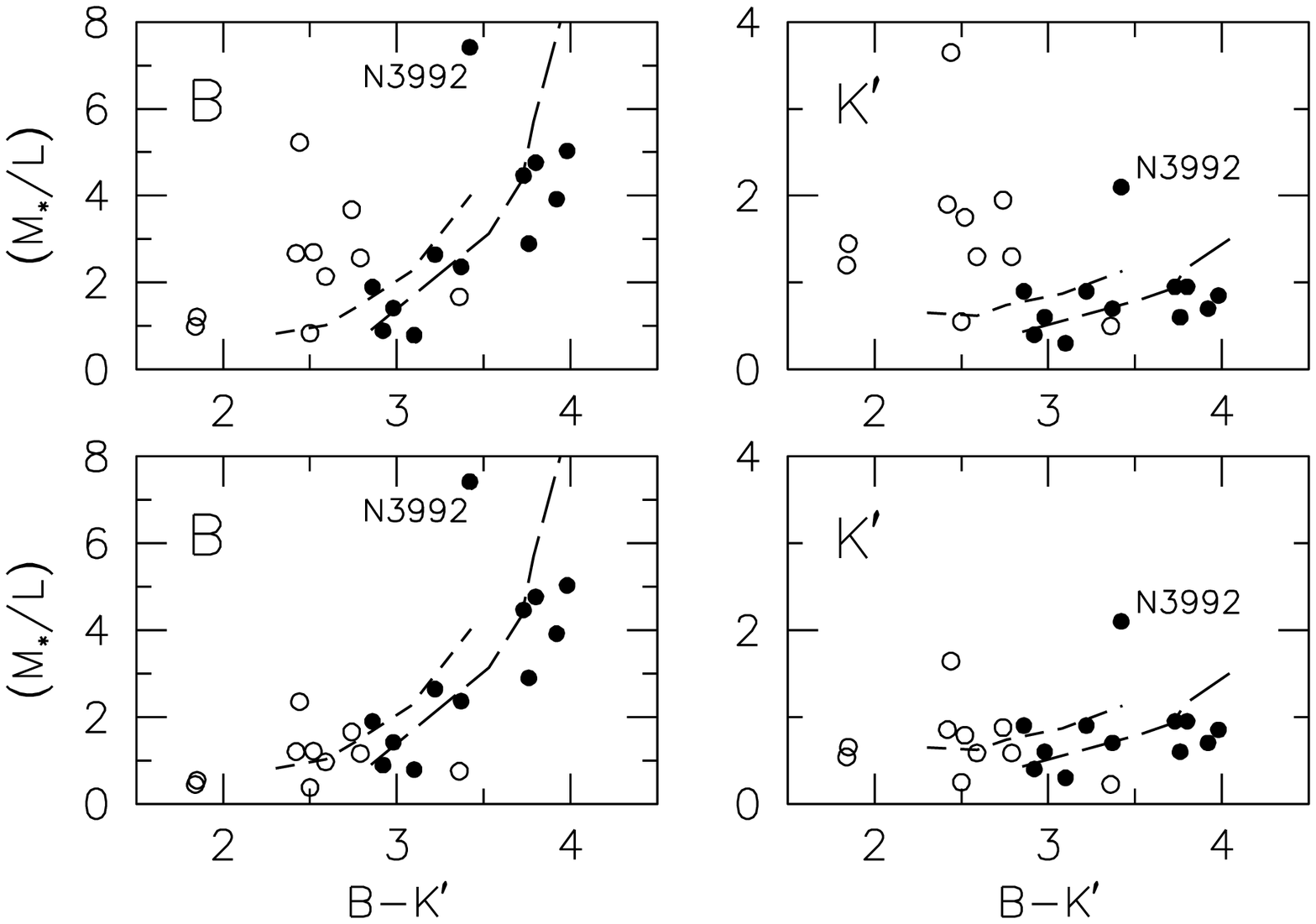} 
\caption{Stellar mass-to-light ratios in the B- and K$^\prime$-band as a
function of B-K$^\prime$ color for HSB (filled symbols) and LSB galaxies
(open symbols).  {\bf Upper panels:} Mass-to-light ratios as derived
from maximum-disk fits for both HSB and LSB galaxies.  {\bf Lower
panels:} Same as upper panels for the HSB galaxies but the
K$^\prime$-band mass-to-light ratios of the LSB galaxies were scaled
down such that
$<$M$_*$/L$_{\rm K^\prime}$$>$$_{\rm LSB}$~=~$<$M$_*$/L$_{\rm
K^\prime}$$>$$_{\rm HSB}$~=~0.71. 
The dashed lines indicate expectations from the Bruzual and Charlot
(1993) stellar population models with solar metallicities, a Salpeter
IMF and ages of 2$-$17 Gyr.  Long dashes: single burst, short dashes:
exponential SFR. }
\end{figure}

\section{Stellar Mass-to-Light Ratios}

We derived extended rotation curves of 22 HI-rich, non-interacting and
sufficiently inclined galaxies in the complete sample which allow us to
investigate a possible relation between this Near-Infrared surface
brightness bimodality and the mass-to-light ratios of the stellar
populations and kinematics of the stellar disks.  The HI rotation curves
were decomposed using the K$^\prime$ luminosity profiles, the HI surface
density profiles and an isothermal sphere model for the dark matter halo
(Figure 5).  Maximum-disk fits were made, avoiding a hollow halo core. 
The resulting stellar mass-to-light ratios in the K$^\prime$-band are
plotted versus B$-$K$^\prime$ color in the upper panels of Figure 4. 

First, let's consider the results for the HSB galaxies (filled symbols). 
In the K$^\prime$-band, the scatter in the stellar mass-to-light ratios
of the HSB galaxies is small (0.22) and the mean value of
$<$M$_*$/L$_{\rm K^\prime}>_{\rm HSB}$=0.71 (excluding NGC 3992) is in
accordance with stellar population models (dashed lines) by Bruzual and
Charlot (1993).  Furthermore, for the HSB galaxies, there is a clear
trend of the B-band mass-to-light ratios with B-K$^\prime$ color as
would be expected from stellar population models.  From a stellar
population point-of-view the maximum-disk hypothesis seems to make sense
for HSB galaxies.  Note that if HSB galaxies are systematically
half-maximum-disk, the small K$^\prime$-band scatter would remain but
the average value would drop to $<$M$_*$/L$_{\rm K^\prime}>_{\rm HSB}$=0.18 and
the trend of (M$_*$/L$_{\rm B}$) with B-K$^\prime$ color would become too
shallow. 

The LSB galaxies on the other hand show a much larger scatter, have a
significantly higher average mass-to-light ratio in the K$^\prime$-band
and do not show any trend of (M$_*$/L$_{\rm B}$) with B-K$^\prime$ color. 
From this we conclude that we overestimate the mass-to-light ratios for
LSB galaxies and, consequently, that it is unlikely that LSB galaxies in
general are in a maximum-disk situation.  This is also supported by the
fact that the higher mass-to-light ratios for LSB galaxies can be
understood easily from a geometrical point-of-view as explained by Zwaan
{\it et al} (1995). 

The lower panels of Figure 4 show the mass-to-light ratios in case the
HSB galaxies have a maximum-disk (same as upper panels) but the
K$^\prime$-band mass-to-light ratios of the LSB galaxies are scaled down
such that the average mass-to-light ratios of HSB and LSB galaxies are
the same;
$<$M$_*$/L$_{\rm K^\prime}$$>$$_{\rm LSB}$~= $<$M$_*$/L$_{\rm
K^\prime}$$>$$_{\rm HSB}$~=
0.71.  Although the scatter in (M$_*$/L$_{\rm K^\prime}$) of the LSBs is
still somewhat larger than that of the HSB galaxies, the trend of
(M$_*$/L$_{\rm B}$) with B-K$^\prime$ color seems to be continued by the LSB
galaxies in the lower left panel.  The luminosities of both HSB and LSB
galaxies were corrected for internal extinction in similar ways. 

\begin{figure}
\plotone{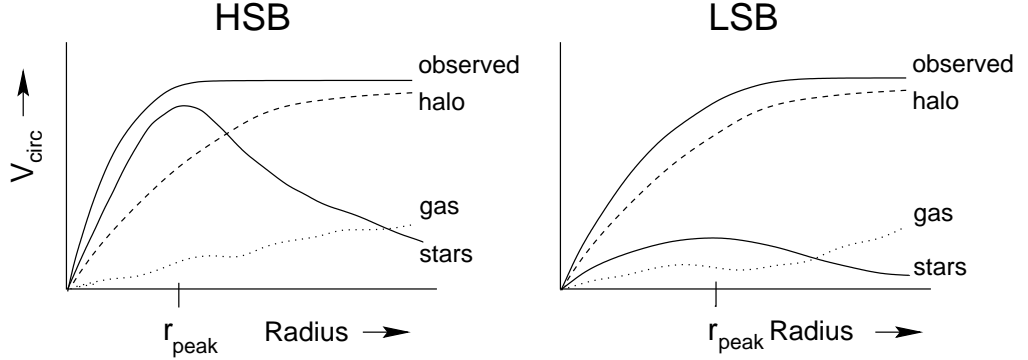}
\caption{Schematic difference between the kinematics of an HSB and an
LSB galaxy with a similar observed maximum rotational velocity.}
\end{figure}

\section{A Dynamical Instability}

Next, we investigate the relative dynamical importance of the dark
matter halo, assuming equal stellar mass-to-light ratios in the
K$^\prime$-band for both HSB and LSB galaxies.  Again, we decomposed the
HI rotation curves but this time we assumed (M$_*$/L$_{\rm K^\prime}$)=0.6
for all galaxies.  This mass-to-light ratio is slightly less than the
average in order not to violate the maximum-disk situation too badly for
most systems.  We calculated for each stellar disk the maximum
rotational velocity V$^{\rm max}_{\rm disk}$ which occurs at r$_{\rm peak}$ around
2.1 disk scale lengths and the rotational velocity
V$_{\rm halo}$(r$_{\rm peak}$) induced by the dark matter halo, at the same
radius where the rotation curve of the stellar disk peaks.  If
V$^{\rm max}_{\rm disk}$$>$V$_{\rm halo}$(r$_{\rm peak}$) then the stellar disk
dominates the potential in the inner regions while the dark matter
dominates if V$^{\rm max}_{\rm disk}$$<$V$_{\rm halo}$(r$_{\rm peak}$).  The
contributions by the gas component to the potential have been ignored. 
Figure 6 shows the measured ratios
V$^{\rm max}_{\rm disk}$/V$_{\rm halo}$(r$_{\rm peak}$). 

Again, we find a bimodal distribution along the lines of the HSB/LSB
dichotomy; HSB disks are self-gravitating and close to a maximum-disk
situation while the potentials of LSB galaxies are dominated by the dark
matter halo at all radii.  Furthermore, it seems that stellar disks
avoid a situation in which the disk is marginally self-gravitating
V$^{\rm max}_{\rm disk}$$\approx$V$_{\rm halo}$(r$_{\rm peak}$).  This
seems to hint at an instability, either in the dynamics of an
established dissipationless stellar disk or during the formation process
of a galaxy when it consisted mostly of dissipational gas.  Testing the
hypothesis of a dynamical instability in a marginally self-gravitating
baryonic disk would require N-body (hydro-)dynamical simulations at a
very high dynamic range.

\begin{figure}
\plotone{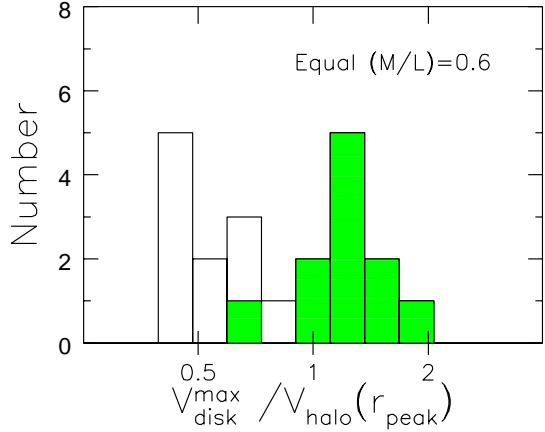}
\caption{Ratio of the maximum rotational velocity induced by the stellar
disk V$_{\rm disk}^{\rm max}$ over the rotational velocity induced by the halo
at r$_{\rm peak}$ where V$_{\rm disk}^{\rm max}$ occurs. The filled histogram is for
HSB galaxies, the open histogram for LSB galaxies. An equal stellar
mass-to-light ratio in the K$^\prime$-band of 0.6 is assumed for all
galaxies. NGC 3992 is omitted.}
\end{figure}

\section{Summary}

Our findings and conclusions can be summarized as follows.

\begin{itemize}

\item There is an avoidance of a domain ($\mu_0^{\rm
i}$(K$^\prime$)$\approx$18.5 mag/arcsec$^2$) of Near-Infrared disk
central surface brightness.  For isolated galaxies, there is a distinct
gap (factor 10 in density) between HSB and LSB families. 

\item The Luminosity Function of the HSB family is essentially truncated
faintward of M$_{\rm B}$=$-$17$^{\rm m}$.  The LSB family Luminosity
Function is sharply increasing at our completion limit of M$_{\rm
B}$=$-$16.5$^{\rm m}$.  There might be a shallow dip in the
K$^\prime$-band Luminosity Function around M$_{\rm
K^\prime}$=$-$21.5$^{\rm m}$. 

\item Although HSB and LSB galaxies with similar (L,V$_{\rm max}$) lie
at the same position on the luminosity-line width relation, they may lie
in very different domains in surface brightness-scale lengths plots. 

\item Maximum-disk decompositions of HSB rotation curves give reasonable
Near-Infrared stellar mass-to-light ratios (average of 0.7 with rms
scatter of only 0.2) and an obvious trend of B-band mass-to-light ratio
with B-K$^\prime$ color.  Maximum-disk decompositions of LSB galaxies
give mass-to-light ratios with a higher average, a larger scatter and no
trend with color.  We infer that HSB disks are close to maximum-disk and
self-gravitating inside $\sim$2 scale lengths while LSB galaxies are
well below maximum-disk and dark matter dominated at all radii. 

\item Assuming that both HSB and LSB galaxies have similar Near-Infrared
mass-to-light ratios, we infer that galaxies avoid a situation in which
their baryonic matter is marginally self-gravitating. 

\item The V$^{\rm max}$ that is relevant for the luminosity-line width
relation is given by the mass distribution of the dark matter halo while
the total luminosity of the dissipational matter is in fixed proportion
to the halo mass.  The exact distribution of the luminous matter within
the dark matter halo is irrelevant for the luminosity-line width
relation.  In high angular momentum LSB systems, the dissipational
matter reaches its rotational equilibrium at large radii and the mass
distribution of dark matter remains dominant.  In low angular momentum
HSB systems, the dissipational matter collapses enough to become
self-gravitating and might even give rise to a rotation curve which is
declining in the inner parts before it becomes flat in the dark matter
supported outer regions. 

\end{itemize}

\section{Future work}

One might argue that our statistics are not overwhelming or that our
findings are particular to the Ursa Major cluster.  To tackle both
issues we are currently analyzing the results of a blind VLA survey in
the Perseus-Pisces ridge which yields an HI selected and volume limited
sample of galaxies of both high and low surface brightness.  The volume
surveyed is similar to the entire Ursa Major volume.  Near-Infrared
surface photometry is planned for the fall of 1999 which should give us
the Near-Infrared surface brightness distribution of this HI selected
sample.  In this way we hope to avoid any optical selection effects that
might be strongly biased in favor of HSB systems.  The combined data
sets will yield a sample that is twice as large as the current Ursa
Major sample itself. 

Of course, we eagerly await the results of any other ongoing project
aimed at measuring the {\it Near-Infrared} surface brightness
distribution. 

\acknowledgments

Mike Pierce, Jia-Sheng Huang and Richard Wainscoat participated in the
collection of the data.  The Westerbork Synthesis Radio Telescope is
operated by the Netherlands Foundation for Research in Astronomy, with
financial support by the Netherlands Organization for Scientific
Research (NWO).  This research has been supported by NATO Collaborative
Research Grant 940271 and grants from the US National Science
Foundation.

\end{document}